\documentclass[twocolumn,secnumarabic,amssymb, nobibnotes, aps, prd]{revtex4}

\usepackage{color}
\usepackage{pstricks}

\begin{document}

\title{Electronic  zero modes of vortices in Hall states of gapped graphene}

\author{Gordon W. Semenoff}
\affiliation{
Department of Physics and
Astronomy, University of British Columbia\\6224 Agricultural Road,
Vancouver, British Columbia V6T 1Z1}
\begin{abstract}
Recent observation of a metal-insulator phase transition in the $\nu=0$ Hall state of graphene has
inspired the idea that charge carriers in the metallic state could be fractionally charged vortices.
We examine the question of whether vortices in particular gapped states of graphene and subject to external
magnetic and pseudo-magnetic fields could have
the mid-gap zero mode electron states which would allow them to be charged.
\end{abstract}
\maketitle

\section{Introduction}

It is now established that, at sufficiently low temperature and strong magnetic field,
the four-fold degenerate Landau level which resides at
the apex of the Dirac cone in graphene is split into four sublevels
\cite{zhang}-\cite{ong1}.  At least some of this splitting is
attributed to spontaneous breaking of the sublattice symmetry and
generation of a mass gap in the electron spectrum.  The further recent
observation of Kosterlitz-Thouless scaling of the resistivity near the
metal-insulator transition which occurs as the strength of the field
is varied \cite{ong1}-\cite{zhang2010} has inspired a number of
possible explanations, perhaps the most interesting of which has
fractionally charged vortices as the carriers of electric charge in
the metallic phase\cite{Nomura}\cite{Hou2009}\cite{Herbut:2009xt}.
This scenario would give another concrete physical
example beyond polyacetylene~\cite{Su:1979ua} of topological solitons
 with fractional quantum numbers.

The vortices are topological defects in the condensate which creates the
mass gap.
In a system with particle-hole symmetry, electron energy
states above and below the middle of the mass gap are paired and
fractional charge can only arise from unpaired mid-gap
states\cite{Jackiw:1975fn}.
The possibility that the interaction of vortices with
relativistic fermions can result in mid-gap states has
been known for a long time
\cite{Jackiw:1981ee}-\cite{Niemi:1984vz}. The novel feature of the
Hall system is the presence of a macroscopic magnetic field where the electron
spectrum would normally be concentrated in Landau levels.
In the absence of the mass condensate, or other symmetry breaking,
a four-fold degenerate (two spins and two valleys)
Landau level
resides precisely at the Dirac point \cite{Semenoff:1984dq}.  With a
parity and time-reversal invariant
mass gap $2m$, this Landau level is split into two levels which
are displaced to the positive
and negative mass thresholds, $E=\pm m$
\cite{Niemi:1984vz} and, without defects, the mass gap contains no
electronic states at all.  It is then reasonable to ask whether a
vortex defect in the condensate which creates the mass gap
can bind a mid-gap electronic state.  This
question has been studied recently by Hou et.al.~\cite{Hou2009} and Herbut \cite{Herbut:2009xt} who
found, surprisingly, that the answer is yes.  In fact, the
existence of mid-gap states attached to vortices is relatively
insensitive to the presence of a magnetic field. In the following,
we shall examine this phenomenon
in more detail.  We will show that the fermion-vortex system in
simultaneous magnetic and pseudo-magnetic fields generally has
$|n|$ zero modes, where $n$ is the vorticity,
even when the magnetic field is a constant so that its
flux diverges with the volume of the system.  The
exception is where the pseudo-magnetic flux inside a disc of large radius $r\to\infty$
grows faster than $r$ and, in addition, when it is larger than the magnetic flux
at large $r$.
In that case, the number of zero modes no longer depends on the vorticity, instead
there are two infinitely degenerate Landau
levels at zero energy.

We shall also show that, when the pseudo-magnetic field falls off at least
as fast at $1/r^2$ at $r\to\infty$, an external magnetic field has no effect on  the
$\eta$-invariant of the graphene Dirac operator perturbed by a
constant mass term which upsets the particle-hole symmetry.
$\eta$ is a topological invariant and it is of physical interest because it
is proportional to the electric charge of a system of electrons which is
governed by the single-particle Hamiltonian in question and in a state where all of the negative
energy levels are filled and all of the positive energy levels are empty.
This result indicates that the fractional charge of a vortex is indeed a topological
invariant and that it is independent of the external magnetic field.

Finally, we shall find interesting correlations between the sublattice components of
the electron mid-gap state wavefunctions and its internal quantum numbers, such as spin or valley assignment.
These correlations are analogous to the Dirac point Landau level in ungapped graphene,
where, for one valley the Landau level zero mode wavefunctions all live on one of the two sublattices and
the other valley zero mode wavefunctions live on the other sublattice, which sublattice depends on the
sign of the external magnetic field.  Zero modes in a vortex background
also follow a pattern that is similar to this.  We will elaborate in later Sections.

A number of scenarios for mass gap generation have been proposed
\cite{Nomura2}-\cite{hatsugai}.  To support topological vortices, the
mass condensate should have an SO(2) symmetry.  Two proposals have
appeared in recent literature.  One is antiferromagnetic  order
with an easy plane \cite{Herbut:2007zz}.  Formation of antiferromagnetic order
would be a result of the short-ranged on-site Coulomb repulsion of
electrons whose role is enhanced in a strong magnetic field\cite{catalysis}.
The Zeeman coupling of electron spins to the magnetic
field would result in an ``easy plane'' where the antiferromagnetic order
parameter lies in the plane that is perpendicular to the applied field.
This is easy to demonstrate using second order degenerate perturbation
theory with the Zeeman coupling as a perturbation and the possible
orientations of the antiferromagnetic condensate as the set of degenerate
ground states.
The other scenario with SO(2) symmetry is
a bond density wave called the Kekule distortion which is a result of
the electron-phonon
coupling\cite{Hou:2006qc},\cite{Nomura}\cite{Jackiw:2007rr}-\cite{Chamon:2007hx}.
It was shown to have vortex defects which, when coupled to the
graphene electron, bind an unpaired mid-gap state.

\section{Graphene Hamiltonian with a mass gap}

We shall begin by considering graphene at energies near the Dirac
points, and with a mass generated by antiferromagnetic order.
The bipartite honeycomb lattice of graphene  is a superposition of two triangular
sublattices.
Antiferromagnetic order gives each of the two spin
states of the electron a charge density wave -- spin up electrons have
higher density on one of the two sublattices and spin down electrons
on the other sublattice.  This breaks the symmetry under interchanging
the two sublattices and
the charge density wave generates a parity-invariant mass term in graphene.~\cite{Semenoff:1984dq}.
The electrons are described by the Hamiltonian~
\begin{equation}\label{hamiltonianwithspincondensate2}
H=\left[ \begin{matrix}{ i\hbar v_F\vec\sigma\cdot\vec D+
\sigma^3\vec m\cdot\vec\tau & 0 \cr
0 & -i\hbar v_F\vec\sigma^*\cdot\vec D+\sigma^3\vec m\cdot\vec\tau\cr
 }\end{matrix}\right]
\end{equation}
This Hamiltonian is an $8\times8$ matrix.
The $2\times2$ structure which is displayed explicitly refers to the two
graphene valleys. The four additional dimensions which are suppressed
are a direct product of two dimensional matrices: one on which
is described by the pseudo-spin Pauli matrices
$\sigma^a$ act and one by the spin Pauli matrices $\tau^a$.
An alternative way of presenting this Hamiltonian is by writing the $8\times8$
matrices as ordered direct products of the three $2\times2$  matrices with
unit and Pauli matrices for pseudo-spin $({\cal I},\sigma^a)$,
spin $({\cal I},\tau^b)$ and  valleys $({\cal I},\eta^c)$, in that order.  In terms of these matrices, (\ref{hamiltonianwithspincondensate2})
has the form
\begin{eqnarray}\label{hamiltonianwithspincondensate3}
H=i\sigma^1\otimes{\cal I}\otimes\eta^3~D_1
+i\sigma^2\otimes{\cal I}\otimes{\cal I}~ D_2 \nonumber \\
+\sigma^3\otimes\tau^a\otimes{\cal I}~~m^a
\end{eqnarray}
 Here, the covariant derivative is $\vec D =
\vec\nabla - i\frac{e}{c}\vec A$ with $\vec A$ the electromagnetic vector
potential and $v_F\approx\frac{c}{300}$ is the graphene
Fermi velocity. The magnetic field is $B_A\equiv \vec\nabla\times\vec A$.

 It is useful to remind the reader that,
 the upper component of the spinors on which the pseudo-spin matrices
 $({\cal I},\sigma^a)$ act has support on one of the graphene
 sublattices, the lower component on the other.

 In (\ref{hamiltonianwithspincondensate2}) and (\ref{hamiltonianwithspincondensate3}),
the masses of each spin polarization are $\pm|m|$, the eigenvalues of
$m_a\tau^a$.
If the antiferromagnetism has an ``easy plane'', only
two of the the three spin matrices can be used to describe possible
orientations of the condensate, for example, $\vec m\cdot\vec \tau =
m_1\tau^1+m_2\tau^2$.  We will be interested in the circumstance where
$m_a$ can depend on spatial position and can have a vortex profile,
with asymptotic limit
\begin{equation}\label{massasymptotic}
\lim_{r\to\infty}\left[m_1(r,\theta)+im_2(r,\theta)\right]
=\hat me^{in\theta}+{\cal O}\left(\frac{1}{r}\right)
\end{equation}
with $(r,\theta)$ the polar coordinates of
the plane and $\hat m$  a constant.

If the mass term in (\ref{hamiltonianwithspincondensate2}) were
absent, the existence of two valleys and two spin states give the
Hamiltonian an SU(4) symmetry. The mass term reduces this to an
$SU(2)$ symmetry which mixes the valleys.\footnote{If the condensate $m_a(x)$ were
a constant, the residual symmetry would be $SU(2)\times SU(2)$.}
It is generated by
\begin{equation}\label{su2}
{\cal I}\otimes{\cal I}\otimes\eta^3~,~
\sigma^2\otimes\tau^3\otimes\eta^1~,~
\sigma^2\otimes\tau^3\otimes\eta^2
\end{equation}

Before we proceed, we observe that,  we can redefine the matrix components
of the Hamiltonian (\ref{hamiltonianwithspincondensate3}) by conjugating it with the matrix
$$
{\cal I}\otimes{\cal I}\otimes\frac{1}{2}({\cal I}+\eta^3)
+\sigma^2\otimes\tau^3\otimes\frac{1}{2}({\cal I}-\eta^3)
$$
then interchanging the   spin and valley labels, $\tau^a\leftrightarrow\eta^a$ and
then conjugating with
$$
{\cal I}\otimes{\cal I}\otimes\frac{1}{2}({\cal I}+\eta^3)
+\sigma^3\otimes{\cal I}\otimes\frac{1}{2}({\cal I}-\eta^3)
$$ the resulting Hamiltonian
is identical to the
one  with a mass term arising from the Kekule distortion that
was discussed in \cite{Hou:2006qc} and \cite{Jackiw:2007rr}.  This replacement amounts to
an SU(4) rotation and other orientations of the SU(4) breaking terms in the Hamiltonian are
possible as well.  We also recall
that in Ref.\cite{Jackiw:2007rr}, when the mass is oriented as coming from a Kekule
distortion, they add a pseudo-magnetic
gauge field which couples to the Kekule condensate.
In the present case where the condensate is antiferromagnetic order, the
analogous gauge field would modify the covariant derivative as $\vec\nabla-i\frac{e}{c}\vec A \to
\vec\nabla-i\frac{e}{c}\vec A -i\vec V\tau^3$ and it would add a term $$\sigma^1\otimes\tau^3\otimes
\eta^3~V_1 + \sigma^2\otimes\tau^3\otimes
{\cal I}~V_2$$ in (\ref{hamiltonianwithspincondensate3}).  In the following, we shall
include this gauge field in our analysis.  In the case where the
condensate is antiferromagnetic, it amounts to gauging the symmetry that
rotates the antiferromagnetic order parameter, as might be done to
describe a spin liquid.

\section{Zero modes}

Since the valley degrees of freedom in (\ref{hamiltonianwithspincondensate2})
are decoupled, we can perform our analysis for each
valley separately.  Consider the Hamiltonian describing
electrons in one of the valleys (for simplicity, in units where  $\hbar = v_f=\frac{e}{c}=1$),
\begin{equation}\label{valleyhamiltonian}
 h= i\sigma^i\otimes{\cal I}\nabla_i + \sigma^i\otimes{\cal I} A_i +\sigma^i\otimes\tau^3 V_i +
\sigma^3\otimes\tau^a~m_a
\end{equation}
As we have suppressed the valley degree of freedom ($\eta^a$ are missing), this is a $4\times4$ matrix differential
operator.

We will search for  solutions of  $h\psi_0=0$. We note that the Hamiltonian (\ref{valleyhamiltonian})
anticommutes with the matrix $\sigma^3\otimes\tau^3$,
\begin{equation}\label{anticommutes}
\sigma^3\otimes\tau^3~ h~ +~ h ~\sigma^3\otimes\tau^3 = 0
\end{equation}
Zero modes of $h$ can also be chosen to be eigenvectors of $\sigma^3\otimes\tau^3$ and we shall show below
that all of the zero modes are indeed eigenvectors with eigenvalues $\sigma^3\otimes\tau^3=-{\rm sign}(m)$.
This  implies that the sublattice position and the spin orientation of the zero mode
are correlated. One sublattice has $\sigma^3=1$ and on that sublattice the spin is polarized as $\tau^3={\rm sign}(n)$.
The other sublattice has $\sigma^3=-1$ and the opposite spin polarization $\tau^3=-{\rm sign}(n)$.  Furthermore, we shall
find that, close to the vortex core, the zero modes have support on both sublattices.  On the other hand, far from the vortex core, the
zero modes have support on only one of the sublattices and therefore have one particular orientation of the spin.
Which of the two
possible sublattices is
chosen depends on the orientation of the external magnetic field.   Also, we note that the spin
polarizations that we are discussing here are always orthogonal to the
spin polarization of the condensate which is in the $\tau^1$-$\tau^2$-plane.

Also note that, if we include  the second valley, its Hamiltonian differs from (\ref{valleyhamiltonian}) by conjugation with the
matrix ${\cal I}\otimes\sigma^2\otimes\tau^3$ which interchanges the sublattices.  Thus, for the other
valley, there are also $|n|$ zero modes.  They also have a peculiar spin polarization to sublattice relationship
which is precisely the opposite of the one in the first valley.

We will begin with the special case of
a rotationally covariant vortex with vorticity $n$:
$(m_1,m_2)=m(r)(\cos n\theta,\sin n\theta)$ where $m(r)\to\hat m$ at
$r\to\infty$ as in (\ref{massasymptotic}) and $m(0)=0$.
We shall also assume
that the magnetic fields $B_A $ and $B_V=\vec\nabla\times\vec V$
depend only on $r$.
We will use the gauge $A_r=0$ and $V_r=0$ where $A_\theta(r)=\int_0^r r'dr'
B_A(r')$ and $V_\theta(r)=\int_0^r r'dr'
B_V(r')$. For example, if the magnetic fields are constants, the vector potentials would be
$A_\theta(r )= \frac{B_A}{2} r^2$ and $V_\theta(r )=
\frac{B_V}{2} r^2$.
When written in polar coordinates, the Dirac equation for a zero mode is
\begin{eqnarray}
m(r)e^{in\tau^3\theta}\tau^1u +
ie^{-i\theta}\left(\partial_r - \frac{i}{r}\partial_\theta
-\frac{A_\theta+\tau^3 V_\theta}{r}\right)v=0
\label{de11}
\\
 ie^{i\theta}\left(\partial_r + \frac{i}{r}\partial_\theta+\frac{A_\theta
+V_\theta\tau^3}{r}\right)u-m(r)e^{in\tau^3\theta}\tau^1v=0
\label{de22}
\end{eqnarray}
The spinor components $u$ and $v$ are eigenstates of pseudo-spin $\sigma^3\otimes{\cal I}$ with eigenvalues $+1$ and $-1$,
respectively.   Formally, they can be defined as
$$
u=\frac{1}{2}\left({\cal I}+\sigma^3\right)\otimes {\cal I}~\psi_0
~~,~~
v=\frac{1}{2}\left({\cal I}-\sigma^3\right)\otimes {\cal I}~\psi_0
$$
These project two two-component spinors,
 $u$ and $v$, from the  four-component spinor $\psi_0$.
$u$ and $v$   each have two spin components on which the Pauli matrices $\tau^a$ act.
We make the Ans\"atze
\begin{eqnarray}
u(r,\theta)&=& e^{i\ell\theta}~\tau^1~\tilde u(r)  \\
v(r,\theta)&=&e^{-ik\theta}\tilde v(r)
\end{eqnarray}
Plugging this into (\ref{de11}) and (\ref{de22}) yields
\begin{eqnarray}
m(r)e^{i(\ell+k +n\tau^3+1)\theta}\tilde u +
i\left(\partial_r
-\frac{k+A_\theta+\tau^3 V_\theta}{r}\right)\tilde v=0
\nonumber
\\
 ie^{i(\ell+k +n\tau^3+1)\theta}\left(\partial_r -\frac{\ell-A_\theta
+V_\theta\tau^3}{r}\right)u-m(r)\tilde v=0
\nonumber
\end{eqnarray}
Now, we shall have to choose the spinors $\tilde  u$ and $\tilde v$ to be eigenvectors of
$ \tau^3$, both with the same eigenvalue $-{\rm sign}(n)$,
$$
\tau^3\tilde u =-{\rm sigma}(n)\tilde u ~~,~~\tau^3 \tilde v =-{\rm sign}(n)\tilde v
$$
(The reason for
this choice of sign will become clear when we find the possible values of $k$ and $\ell$
which both will have to be positive and have to obey (\ref{kandell}) below.)
$\tilde u$ and $\tilde v$ now ech have only one non-zero component.

Then, the $\theta$-dependent
phases cancel if
\begin{equation}\label{kandell}
 \ell+k =|n|-1
\end{equation}
which we shall assume holds. Consistency of (\ref{kandell}) is a result of the correct choice
of sign of the eigenvalues of $ \tau^3$.

What remains is
\begin{eqnarray}
m(r)  \tilde u + i \left(\partial_r
-\frac{k+A_\theta -{\rm sign}(m) V_\theta}{r}\right)\tilde v=0
\label{de1}
\\
 i\left(\partial_r -\frac{\ell - A_\theta
-{\rm sign}(m)V_\theta}{r}\right) \tilde u-m(r)\tilde v=0
\label{de2}
\end{eqnarray}

Assuming that $A_\theta$, $V_\theta$ and $m(r)$ go to zero at small $r$,
by examining (\ref{de1}) and (\ref{de2}) at $r\sim 0$,
we see that the small $r$ behavior of the wave-functions must be
\begin{eqnarray}
 \tilde u(r)&\sim& r^{k}\tilde u_0
\label{de5}
\\
\tilde v(r)&\sim &r^\ell\tilde v_0
\label{de6}
\end{eqnarray}
We shall need both of these solutions in order to match the solution at
$r\to\infty$ where there will
generally only be one solution.  Both solutions are normalizable at $r=0$ if
$\ell=0,1,2,...$ and if $k=0,1,2,...$.  However, remembering that $k$ and $\ell$ are related by
(\ref{kandell}), we see that the allowed  solutions  are
\begin{equation}\label{rangeofell}
 (\ell,k) = (0,|n|-1),(1,|n|-2),...,(|n|-1,0) ~,~|n|\geq 1
\end{equation}
We conclude that there are exactly $|n|$ possible solutions where both
behaviors are normalizable at the origin.

The spinor components $\tilde u$ and $\tilde v$ are
then both eigenvectors of spin ${\cal I}\otimes\tau^3$ with the same eigenvalue:
${\cal I}\otimes\tau^3\tilde u=  -{\rm sign}(n)\tilde u$  and ${\cal I}\otimes\tau^3\tilde v=  -{\rm sign}(n)\tilde v$.
Recalling that $u$ and $\tilde u$ differ by a factor of ${\cal I}\otimes\tau^1$ which flips the sign
of the eigenvalue of ${\cal I}\otimes\tau^3$, we conclude that $u$ and $v$ are eigenvectors of spin with
\begin{eqnarray}
{\cal I}\otimes\tau^3 u&=&+{\rm sign}(n)u \\
{\cal I}\otimes\tau^3 v&=& -{\rm sign}(n)v
\end{eqnarray}
Also, we recall that $u$ and $v$ are eigenvectors of pseudo-spin $\sigma^3\otimes{\cal I}$:
\begin{eqnarray}
\sigma^3\otimes{\cal I}u~=~  +~u \\ \sigma^3\otimes{\cal I}v~= ~ -~v
\end{eqnarray}  Combining the
eigenvalues tells us that $u$ and $v$ are
eigenvectors of  $\sigma^3\otimes\tau^3$ with
the same eigenvalue, ${\rm sign}(n)$:
\begin{eqnarray}
\sigma^3\otimes\tau^3u&=& {\rm sign}(m)u \\
\sigma^3\otimes\tau^3v&=& {\rm sign}(m)v
\end{eqnarray}
or, more succinctly
\begin{equation}
\sigma^3\otimes\tau^3\psi_0= {\rm sign}(m)\psi_0
\end{equation}
Remember
that $\sigma^3\otimes\tau^3$ anticommutes with the Hamiltonian, so we expect that the
zero mode wave-functions $\psi_0$ are are also eigenvectors of this matrix.
Here, we see explicitly that is is the
case and the eigenvalue is determined by the sign of the vorticity. This establishes
the basis for the discussion after equation (\ref{anticommutes}) above.

Now we shall consider the large $r$ limit of (\ref{de1}) and
(\ref{de1}).  If $A_\theta$ and $V_\theta$ grow slower than $r$ at
$r\to\infty$ (meaning that $B_A$ and $B_V$ fall off faster than $1/r$),
the solution at large $r$ is identical to the one found by Jackiw
and Rossi \cite{Jackiw:1981ee} and their conclusion that there are exactly $|n|$ zero modes
applies here.
To generalize this, we
will assume that at least one of $A_\theta$ and $V_\theta$ grow faster
than $r$ as $r\to\infty$. (This means that $B_A(r)$ or $B_V(r)$ fall off slower
than $1/r$ and one of both of the fluxes $\int d^2x B_A$ or $\int d^2x B_V$ diverge.)
Then (\ref{de1}) and (\ref{de2}) are solved
at large $r$ by
\begin{eqnarray}
 \tilde u(r)&\sim& e^{\int_0^r\frac{dr'}{r'}(A_\theta(r')
-{\rm sign}(m) V_\theta(r'))} ~\tilde  u_0
\label{de7}
\\
\tilde  v(r)&\sim& e^{\int_0^r\frac{dr'}{r'}(-A_\theta(r') -{\rm sign}(m) V_\theta(r'))}~\tilde v_0
\label{de8}
\end{eqnarray}
We must now choose $\tilde u_0$ and $\tilde v_0$ so that the wavefunctions are normalizable.

 First consider the case where $|A_\theta|>|V_\theta|$. Then, if
$A_\theta>0$ at large $r$, we set $\tilde u_0=0$ and   (\ref{de8}) is
normalizable at $r\to\infty$.  On the other hand, if $A_\theta<0$, we set $\tilde v_0=0$ and
(\ref{de7}) is normalizable. Thus we see that, in
the case where $|A_\theta|>|V_\theta|$, there is always one
normalizable solution of the Dirac equation at large $r$.  Once it is
given the appropriate angle dependence, $e^{i\ell\theta}\tau^1\tilde u(r)$ and
$e^{-ik\theta }\tilde v(r)$ with one of the $|n|$ allowed values of $k$ and
$\ell$, and continued to small $r$, it will become a linear
combination of the two normalizable solutions that we found at $r\sim 0$.  We
conclude that, when $|A_\theta|>|V_\theta|$, there are exactly $|n|$
zero modes, independent of the specific asymptotic behavior of
$A_\theta$.
This is in line with the conclusions in
Refs.~\cite{Hou2009} and \cite{Herbut:2009xt}.

 In addition, in the large $r$ regime,
the solutions reside on either one or the other sublattice (either $u$ or $v$ is non-zero),
depending  on the sign of $A_\theta$ ( and therefore the magnetic flux $\phi_A=\frac{1}{2\pi}\int d^2x B_A$.
Since the spin and the pseudo-spin are
correlated, all zero modes in the asymptotic region
have one particular spin polarization -- the eigenvalue of $\tau^3$ is $-{\rm sign}(\phi_A){\rm sign}(n)$ -- the zero mode in the asymptotic
region is polarized in one particular direction, orthogonal to the easy plane of the condensate. The direction depends
on the sign of the total magnetic flux and the sign of the vorticity.

Now, consider the case where $|A_\theta|<|V_\theta|$.
To understand this case more clearly, we observe that, if $\vec V(x)$ is in the
Coulomb gauge, $\vec\nabla\cdot \vec V=0$, we can rewrite $h$
as
\begin{eqnarray}
 h= e^{\sigma^3\otimes\tau^3\chi_V(x)}
\left\{i\sigma^i\otimes{\cal I}\nabla_i +
 \sigma^i\otimes{\cal I} A_i \right. \nonumber \\ \left. +
\sigma^3\otimes\tau^am_a\right\} e^{\sigma^3\otimes\tau^3\chi_V(x)}
\end{eqnarray}
where $\chi_V(x) = \frac{1}{2\pi}\int d^2x'\ln|\vec x-\vec x'|B_V(x')$.
The zero mode problem for $h$ can be solved by finding potential zero modes when $\vec V=0$
and testing their normalizability at $r=\infty$ when they are multiplied by
the additional factor  $e^{-\sigma^3\otimes\tau^3\chi_V(x)}$. In the case that we
are interested in, when $e^{\pm\chi_V}$ dominates the
asymptotics, there are always two   normalizable
solutions with correlated pseudo-spin and spin so that
$\sigma^3\otimes\tau^3\chi_V(\infty)>0$.  When we begin at small $r$ with one of
the spinors with behavior in (\ref{de5}) or (\ref{de6}),
$\left[\begin{matrix}{e^{i\ell\theta}\tau^1\tilde u(r) \cr 0\cr}\end{matrix}\right]$
or  $\left[\begin{matrix}{0\cr e^{-ik\theta}\tilde v(r) \cr }\end{matrix}\right]$
and extrapolate to large $r$, they should become linear combinations of the two
normalizable solutions there.
In this case, there are an infinite number of zero modes for $\ell=0,1,2,...$ or
$k=0,1,2,...$ which comprise a two-fold degenerate
Landau level at zero energy.  We conclude that,
when $|A_\theta|<|V_\theta|$, there is a Landau level at the Dirac point.
Here, by ``Landau level'' we mean an infinite set of  degenerate states and they exist
whenever $B_V(r)$ falls off slower than $1/r$  at $r\to\infty$.

\section{Index and $\eta$-invariant}

The index of $h$ is the number of zero modes which have eigenvalue
of $\sigma^3\otimes\tau^3=+1$ minus the number with eigenvalue $\sigma^3\otimes\tau^3=-1$.
For generic smooth profiles of $\vec A$, $\vec V$ and $m$,
the index  is a
topological invariant that depends only on the asymptotic values of these functions.
To see this we need only remember that the positive and negative energy states of $h$ are paired:
If $$h\psi_E=E\psi_E$$ then $$h(\sigma^3\otimes\tau^3\psi_E)=-E(\sigma^3\otimes\tau^3\psi_E)$$ and
$$\psi_{-E}=\sigma^3\otimes\tau^3\psi_E$$
If we deform the background fields in $h$ smoothly,
generally the eigenvalues of $h$ move and some modes that are at $E=0$ could move away from $E=0$ and vice versa.  However, since
the non-zero modes come in positive and negative pairs, when modes move to or away from $E=0$, they must do so
in pairs.  Furthermore, these pairs contain exactly one positive and one negative eigenstate of $\sigma^3\otimes\tau^3$:
$$\sigma^3\otimes\tau^3\left(\psi_E+ \psi_{-E}\right)=+ \left(\psi_E+ \psi_{-E}\right)$$ and
$$\sigma^3\otimes\tau^3\left(\psi_E- \psi_{-E}\right)=- \left(\psi_E- \psi_{-E}\right)$$ Thus the difference
between the number of zero modes with $\sigma^3\otimes\tau^3=+1$ and $\sigma^3\otimes\tau^3=-1$, i.e.~the index, is unchanged by a
smooth deformation of the background fields.

In the discussion above, we found that when either the mass term or the magnetic field $B_A$ dominate
the asymptotics, the index is \begin{equation}{\rm Index}(h)=n\label{index1}\end{equation}
On the other hand, when the pseudo-magnetic field $B_V$ is dominant at $r\to\infty$, the index
diverges, as it should equal the total number of states in two infinitely degenerate
Landau levels,   \begin{equation}{\rm Index}(h)\sim   \frac{1}{\pi}\int d^2x B_V(x)\label{index2}\end{equation}
(As we have discussed above, in order for $V_\theta$ to be more important than the mass term at $r\to\infty$,
its fall off must be
slow enough that the flux diverges at least logarithmically in the volume.)
This quantitiy diverges at least logarithmically in the linear size of the 2-dimensional plane
for the case in which it is valid and it would require more care to define it precisely.

Another interesting topological quantity is the $\eta$-invariant,
\begin{equation}\label{001}
\eta\left(h_\epsilon\right) =~{\rm Tr}~{\rm
sign}\left(h_\epsilon\right)
\end{equation}
where
$$
h_\epsilon = h + \sigma^3\otimes\tau^3\epsilon
$$
It is of  interest because the state
of a system of electrons which is governed by the single-particle Hamiltonian $h_\epsilon $ and which has
all negative energy levels occupied by electrons and all positive energy levels empty has electric charge
given by the formula \cite{Niemi:1984vz}
\begin{equation}
\left<Q\right>=\frac{e}{2}\eta(h_\epsilon)
\end{equation}
$\eta$ is formally the difference of two infinite quantities,
the number of states with positive energy and the number of states
with negative energy. In interesting cases, these partially cancel to
leave a finite result.
To make $\eta$ unambiguous, we have added a term to the
Hamiltonian, $h\to h_\epsilon=h+\sigma^3\otimes\tau^3\epsilon$.
The term
$ \sigma^3\otimes\tau^3\epsilon$  is physically
equivalent to an out-of-plane component of the antiferromagnetic order parameter which could be realistic. Since $$\{h,
\sigma^3\otimes\tau^3\}=0$$ the square of $h_\epsilon$ obeys $$h_\epsilon^2=h^2 + \epsilon^2>0$$ and
$h_\epsilon$ has no zero modes.  All of its eigenvalues are either positive or negative. Furthermore, since no matrix
anti-commutes with $h_\epsilon$, the spectrum of $h_\epsilon$ does not
have particle-hole symmetry.  Unpaired zero modes of $h$ are also
eigenfunctions of $h_\epsilon$.  They obey $h\psi_0=0$ and
are eigenvectors of   $\sigma^3\otimes\tau^3$ with
eigenvalues $+ 1$ or $-1$.  They are therefore eigenfunctions
of $h_\epsilon$ with energies $\epsilon$ or $-\epsilon$.
We can recover $h$ by taking $\epsilon$ to zero. Then, the spectrum becomes entirely
symmetric except for unpaired zero modes, which were assigned to the positive
or negative energy states of $h+\sigma^3\otimes\tau^3\epsilon$.

$\eta$ is a topological invariant which is entirely determined by the
asymptotic behavior of $\vec A(x)$, $\vec V(x)$ and $m(x)$. We can easily
derive a formula for $\eta$ when we
assume that the total flux $\phi_V=\frac{1}{2\pi}\int d^2x B_V(x)$ is
finite, but the total electromagnetic flux
$\phi_A=\frac{1}{2\pi}\int d^2x B_A(x)$ need not be.  Computation of
$\eta$ is a straightforward generalization of that
outlined in \cite{Chamon:2007hx} and also
\cite{Niemi:1984gm},\cite{Niemi:1984vz} and will be reviewed in the Appendix.
The result is
\begin{eqnarray}
 \eta(h_\epsilon)=
\left[{\rm sign}(\epsilon)-\frac{\epsilon}{\sqrt{\epsilon^2+\hat m^2}}\right]n
+\frac{2\epsilon}{\sqrt{\epsilon^2+\hat m^2}}\phi_V
\label{result}\end{eqnarray}
To derive this formula, we have had to assume that $\vec V(x)$ is
such that the covariant derivative of the mass term, $\vec\nabla m(x) - i\tau^3\vec V(x)m(x)$
decays at least as fast as $1/r$ at large $r$. (For a vortex solution of
Ginzburg-Landau action, this quantity generally decays exponentially.)

Like the index, $\eta$ in (\ref{result}) does not depend on the electromagnetic gauge field
$\vec A$ at all. There are a few interesting limits of (\ref{result}). If we put $\epsilon\to 0^+$,
we recover the index of $h$ (\ref{index1}). If we put $B_V$ to zero, we see that $\eta$ has an irrational part,
$-\frac{\epsilon}{\sqrt{\epsilon^2+\hat m^2}}n$ which can bee attributed to the polarization of
continuum states by the vortex. Note that this polarization vanishes when $\epsilon\to0$ and
the continuum spectrum becomes symmetric.  The irrational part of the fractional charge of a vortex
in the absence of $\vec V$ is identical to the one noted in
Refs.~\cite{Hou:2006qc},\cite{Chamon:2007pf}-
\cite{Chamon:2007hx},\cite{Hou2009}.
Finally, if we decouple the vortex by setting the mass gap $\hat m\to 0$ we recover (\ref{index2}).

\section{Conclusion}

We have discussed scenarios where vortices can have mid-gap electronic
bound states.  The main conclusion is that such bound states survive, practically unmodified in an external magnetic field.  This in turn
validates the consideration of vortices with mid-gap states as charge carriers in a Hall state.

We comment that, like polyacetylene, where spin degeneracy obscures
fractional charge, so that polyacetylene actually has integer charged
solitons, the system that we have discussed here has a valley degeneracy and the same phenomenon will occur.
A vortex with a zero mode, actually has a pair of zero modes, one for each valley.  Therefore,
instead of being a two level system (zero mode occupied or unoccupied) where the states have charges
$+\frac{1}{2}$ and $-\frac{1}{2}$, it is a four-level system
with charges $-1,0,0,1$.  In polyacetylene, this leads to exotic
spin-charge assignments, the charge $\pm 1$ states of the soliton
are spinless, whereas the two charge $0$ states have spins $\pm
\frac{1}{2}$, a topological version of spin-charge separation.
A similar phenomenon would occur in graphene.  When the mass condensate is
a Kekule distortion, the vortex states would have spin degeneracy and the same
exotic spin-charge assignments as occur in polyacetylene would now appear in two dimensions,
If the condensate is an antiferromagnet, the spin is replaced by the residual SU(2)
valley quantum number, and could be more difficult to see experimentally, as the symmetry
is emergent and the conservation law resulting from it is not exact.

As for conductivity, as in polyacetylene, having vortices with low energy
charged states makes the number of charged vortices available for conduction of electric
current extremely sensitive to any
displacement of the chemical potential from the charge neutral point. A signature of
charged vortices of this kind should be, like in polyacetylene, an enormous variation
of carrier concentration with doping.

\begin{acknowledgements}
 The author acknowledges financial support from NSERC of Canada and the hospitality of the
Aspen Center for Physics where part of this work was done. He also acknowledges Roman Jackiw
for a discussion and some comments on the manuscript.
\end{acknowledgements}

\vskip 1cm
\appendix*
\section{Computation of the $\eta$-invariant}
\vskip .5cm

In this appendix, we outline the computation of $\eta$,
the result of which was quoted in Eq.~(\ref{result}).
We  regulate the trace in (\ref{001}) and represent
the step function  sign($h$) by an integral
\begin{equation}\label{002}
\eta = \frac{2}{\pi} \lim_{\beta\to0}\int_0^\infty
 d\omega ~{\rm Tr}\left[~\frac{h}{h^2+\omega^2}~e^{-\beta h^2}\right]
\end{equation}
which, using the fact that $\{\sigma^3\tau^3,h\}=0$, we replace by
\begin{eqnarray}\label{1019}
\eta=\frac{2\epsilon}{\pi}\lim_{\beta\to0}\int_0^\infty
 d\omega ~
 ~{\rm Tr}\left[~\sigma^3\tau^3\frac{e^{-\beta
h^2}}{h^2+\epsilon^2+\omega^2}~\right]
\end{eqnarray}
We define the current
\begin{equation}
\vec J(x)~=~ {\rm tr}~(x|i\vec\sigma \sigma^3\tau^3
\frac{h_0 e^{-\beta h_0^2}}{h_0^2+\epsilon^2+\omega^2}|x)
\end{equation}
where tr indicates a trace over the Dirac and spin indices
only. (Our previous trace, denoted Tr was a trace over Dirac and
spin indices and position space as well ${\rm Tr} {\cal O} = \int dx~ {\rm
tr~}(x|{\cal O}|x)$.  ) Taking a divergence of the current yields the identity
\begin{eqnarray}\label{351}
\vec\nabla\cdot\vec J&= &
\vec\nabla
\cdot {\rm tr~}(x|i\vec\sigma \sigma^3\tau^3 \frac{h_0e^{-\beta h_0^2}}{h_0^2+
\epsilon ^2+\omega^2}|x)\nonumber \\
&=&{\rm tr~}(x|h_0\sigma^3\tau^3 \frac{h_0e^{-\beta h_0^2}}{h_0^2+
\epsilon ^2+\omega^2}
-\sigma^3\tau^3 \frac{h_0e^{-\beta h_0^2}h_0}{h_0^2+
\epsilon ^2+\omega^2}|x)
\nonumber \\
&=&-2{\rm tr~}(x|  \sigma^3\tau^3  e^{-\beta h_0^2}|x)
\nonumber \\
\label{359}
&&+2(\epsilon ^2+\omega^2)
{\rm tr~}(x|\sigma^3\tau^3 \frac{e^{-\beta h_0^2}}
{h_0^2+\epsilon ^2+\omega^2}|x)
\end{eqnarray}
Using the identity in (\ref{359}), we rewrite (\ref{1019}) as
\begin{eqnarray}
&\eta &=\lim_{\beta\to 0} \int_0^\infty
\frac{d\omega}{\pi}~
\frac{2\epsilon }{\epsilon ^2+\omega^2}\left\{
{\rm Tr}\left[\sigma^3\tau^3  e^{-\beta h_0^2}\right] \right.
\nonumber \\
 &+&\left.\frac{1}{2}
\int d^2x\vec\nabla\cdot
{\rm tr~}(x|i\vec\sigma \sigma^3\tau^3  \frac{h_0e^{-\beta h_0^2}}{h_0^2+
\epsilon^2+\omega^2}|x) \right\}
\label{360}
\end{eqnarray}
The
first term  in (\ref{360})
is  evaluated using an asymptotic expansion of
the heat kernel,
\begin{eqnarray}
 (x|e^{-\beta h_0^2}|x)
=\frac{1}{4\pi\beta}
+\frac{1}{4\pi}\left[\sigma^3(B_A+\tau^3 B_V)  \right. \nonumber \\ \left.
  +\vec\sigma\times\vec Dm -m^2\right]  +{\cal O}(\beta)
\end{eqnarray}
 where $\vec Dm = \left( \vec\nabla -2i\tau^3\vec V\right)m$. When we multiply by $\sigma^3\tau^3$ and
take the trace, we see that the first term in (\ref{360}) is
\begin{equation}\lim_{\beta\to 0} \int_0^\infty
\frac{d\omega}{\pi}~
\frac{2\epsilon }{\epsilon ^2+\omega^2}
{\rm Tr}\left[\sigma^3\tau^3  e^{-\beta h_0^2}\right] =
 {\rm sign}(\epsilon)\frac{1}{\pi}\int d^2x B_V
\end{equation}

Now, we consider the second term on the right-hand-side of (\ref{360}) which,
using Gauss' theorem,  can be written as a surface integral on the
circle at
$r=\infty$.
To evaluate it, we must make some
assumptions about the asymptotic behavior of the fields.   We assume that the gauge field $\vec A$
is such that $B_A $  grows
no faster than a constant at large $r$. This allows the case of a constant magnetic field.  In this case
asymptotically, in a radially symmetric gauge $A_\theta=\frac{B_A}{2}r^2$.
We shall need to assume that the gauge field $\vec V(r)$ decays at least as fast as $1/r$ for large $r$
so that $\vec Dm\sim 1/r$ and $B_V\sim\frac{1}{r^{2 }}$.
We then can use a functional Taylor expansion of the integrand in powers of
$\vec Dm$, where it turns out that only the first (linear) order survives the traces and the asymptotic
and $\beta\to0$ limits. The trace of the zeroth order vanishes. The first and higher order terms in the
expansion are finite in the limit $\beta\to0$ which can now safely be taken.  The first order in $\vec Dm$ is
\begin{eqnarray}
 \lim_{\beta\to0}{\rm tr}(x|i\vec\sigma \sigma^3\tau^3
\frac{h_0 e^{-\beta h_0^2}}{h_0^2+\epsilon^2+\omega^2}|x)= \nonumber
\\
{\rm tr} i\vec\sigma \sigma^3\tau^3\int d^2y ~\sigma^3m(x)g(x,y)
\vec\sigma\times\vec Dm (y)g(y,x)
\label{intstep}\end{eqnarray}
where
\begin{equation}g(x,y)=(x|\frac{1}{
-D^2-\sigma^3 B_A-\sigma^3\tau^3 B_V
  +\hat m^2+\epsilon^2+\omega^2}|y)
\label{greenfunction}\end{equation}
Since the Green functions in (\ref{intstep}) are
short ranged and $|\vec x|\to\infty$, they will have support in the region
where $|\vec y|\to \infty$ and the background fields in (\ref{greenfunction})
can be replaced by their asymptotic values.  Then, the
Green function can be found explicitly,
$$
g(x,y)=\frac{B}{4\pi}\int_0^\infty d\lambda  \frac{ e^{-\chi(x,y,\lambda) }}{\sinh B_A\lambda}
$$
with $\chi(x,y,\lambda) =\frac{B_A}{4}(\vec x-\vec y)^2\coth B_A\lambda
+\lambda(\sigma^3B_A+\hat m^2+\omega^2+\epsilon^2)+i\frac{B_A}{2}\vec x\times\vec y$.
We can now easily show that, because of the large $r$ limit, the green functions
$...g(x,y)...g(y,x)...$ in (\ref{intstep}) can be replaced by
$\frac{1}{4\pi}\frac{1}{\hat m^2+\omega^2+\epsilon^2}\delta^2(x-y)$.
This can be seen by defining $\vec x = r\hat x$, rescaling $\vec y\to ~r~\vec y$
and taking the large $r$ limit.  Then, we have
\begin{eqnarray}
 \lim_{\beta\to 0}{\rm tr}(x|i\vec r\cdot\vec\sigma \sigma^3\tau^3
\frac{h_0 e^{-\beta h_0^2}}{h_0^2+\epsilon^2+\omega^2}|x) \nonumber \\
 =-\frac{1}{\pi}\frac{\epsilon^{ab}m^a{\partial}_{\theta}m^b+2m^2V_\theta }{\hat m^2+\omega^2+\epsilon^2}
\end{eqnarray}
Doing the remaining integrals, we obtain (\ref{result}).

 \end{document}